\documentclass[twocolumn,preprintnumbers,amsmath,amssymb]{revtex4}
\usepackage{amsmath,amssymb,graphics,epsfig,subfigure}
\usepackage{graphicx}
\usepackage{hyperref}
\usepackage{color}
\usepackage{csquotes}
\usepackage{pdfpages}
\usepackage{lipsum}
\usepackage{mathtools}

\begin{document}
	
	\thispagestyle{empty}
	
	\begin{center}
		
		\title{\Large \bf Criticality of charged AdS black holes with string clouds in boundary conformal field theory}

		\author{Aditya Singh$^1,$\footnote{E-mail: 24pr0148@iitism.ac.in} and Sandip Mahish$^2,$\footnote{{{E-mail: sm19@iitbbs.ac.in}} \\ \textbf{\;\;\;\;\;\;\;\;\;\; All Authors Contributed Equally}}}

		\affiliation{ \vspace*{0.1cm} $^1$ Department of Physics, Indian Institute of Technology (Indian School of Mines), Dhanbad, Jharkhand-826004, India \\
		$^2$ Department of Physics, School of Basic Sciences, Indian Institute of Technology, Bhubaneswar, Odisha-752050, India}

\begin{abstract}
The aim of this letter is to study the universal thermodynamics and criticality of charged AdS black holes with string clouds in the bulk and in the boundary conformal field theory (CFT). For this system, we determined the critical quantities and noticed that the free energy in the bulk exhibits swallow tail behavior. In the boundary CFT, the presence of second order phase transition is observed. At constant charge, the heat capacity is finite in the bulk but diverges at critical points in the boundary CFT. Furthermore, we tried to determine the nature of interactions between black hole molecules in the boundary CFT and in the bulk, which is novel for charged AdS black holes with string clouds.
\end{abstract}
\maketitle
\end{center}

\section{Introduction}

Black holes are prevalent topic in discussions of quantum gravity as they are the most promising object to study strong gravity. Since Hawking and Bekenstein's seminal work on black hole thermodynamics\cite{Bekenstein:1972tm,Bekenstein:1973ur,Hawking:1975vcx,Bardeen:1973gs} and the Gauge/gravity duality by Maldacena\cite{Maldacena:1997re}, this topic has taken a novel direction. At large $N$ (number of degrees of freedom),  the relation between strongly coupled field theories in one lower dimension and quantum gravity in AdS spacetime can be explained by gauge/gravity duality; it provides dual desription of quantum gravity. On the other hand, the emergence of quantum gravity can be inferred from black hole thermodynamics. Inspired by the work of Henneaux and Teitelboim\cite{Henneaux:1984ji,Henneaux:1989zc,Dolan:2010ha}, the cosmological constant ($\Lambda$) is regarded as a thermodynamic variable in this formalism. 
\par
Afterwards, Kastor et al.\cite{Kastor:2009wy} proposed that the cosmolgical constant must be varied in order to generalize the Smarr formula from flat black holes to AdS. Pressure $P$ is set as a negative cosmological constant. Small and large black holes phase transition behavior exactly matches with that of liquid gas phase transition analogous to van der waals fluids\cite{Kubiznak:2012wp,Cai:2013qga,Shen:2005nu,Sahay:2010tx,Hawking:1982dh,Chamblin:1999tk}. The degrees of freedom of it's constituents can be inferred from the thermodynamic degrees of freedom of black holes. However, due to black holes intrinsic nature, thermodynamics alone is not sufficient to investigate micromolecules. Weinhold subsequently proposed \enquote{Thermodynamic Geometry} which is a modified version of Einstein's \enquote{Thermodynamic Fluctuation Theory}. This was further developed by Ruppeiner, who examined it for systems, such as van der Waals fluid and ideal gases in\cite{Ruppeiner:1983zz,Ruppeiner:1995zz,Ruppeiner:2007hr}.
\par
It was Letelier\cite{Letelier:1979ej,Chabab:2020ejk} who first introduced a cloud of strings in late 1979 which can be considered analogous to the \enquote*{cloud of dust} in one dimension. A black hole with string cloud resemble purely energetic field configuarations as it's radial configuaration is counter balanced by the external push of a negative pressure applied to the internal pull of the gravitational field\cite{MoraisGraca:2018ofn}. The global monopole in four dimension is one such example. The presence of string cloud acts as an additional gravity source which alters the Schwarzschild solution\cite{Dey:2017xty}, thus enlarging the event horizon radius. One of the major outcome of investigating Einstein's equation with string cloud is that the relativistic strings may be used to define certain suitable models for interactions. Following  the same procedure we will compute thermodynamic scalar curvature for charged AdS black hole with string cloud in $(3+1)$ dimension. Further, depending on the behavior of Ruppeiner scalar curvature we shall investigate the nature of interaction among the micromolecules of black holes. For detailed analysis of thermodynamic geometry of various systems one may see\cite{Janyszek:1989zz,Ruppeiner:1979bcp,Ruppeiner:1981znl,Wei:2019yvs}.
\par
\textbf{Motivations and results:} Strings are considered to be fundamental objects in universe and hence, studying it's effect on various gravitational theory proves to be quite useful. The main motivation of this work is to study how the presence of cosmological cloud of strings in the background of charged AdS black holes would affect the thermodynamics and criticality of the system. It can be noted that these black holes have finite heat capacity in the bulk at constant charge and volume but the heat capacity is divergent at constant charge for the boundary field theory and second order phase transition is observed. Also, pressure jumps across the two phase regime rather than the volume.
\par
This letter is arranged as follows: In section \textrm{II} we will review the basic thermodynamics for charged AdS black holes with string cloud. Section \textrm{III}, discusses the universal thermodynamics and criticality in the boundary CFT. In section \textrm{IV}, we probe thermodynamic geometry and also compute the thermodynamic curvature criticality in the boundary CFT. Finally, we summarize with remarks in section \textrm{V}. 

\section{Charged AdS black hole surrounded with string cloud}
The gravitational action in $(3+1)$ dimension is\cite{Chakrabortty:2011sp},
\begin{equation}
\mathcal{I}_G=\frac{1}{16\pi G_{4}}\int dx^4\sqrt{-g}(R-2\Lambda-\frac{1}{4}F_{\mu\nu}F^{\mu\nu})+\mathcal{I}_{m},
\end{equation}
where the matter part of the action $\mathcal{I}_{m}$ is given as,
\begin{equation}
\mathcal{I}_{m}=-\frac{1}{2}\sum_{i}\mathcal{T}_{i}\int d^2\zeta\sqrt{-h}h^{\alpha\beta}\partial_{\alpha}x^\mu \partial_{\beta}x^{\nu}g_{\mu \nu},
\end{equation}
here, $g_{\mu \nu}$ denotes the spacetime metric, $h^{\alpha \beta}$ is the worldsheet metric\cite{Guo:2025bgx}. $\mathcal{T}_{i}$ represents the tension for $i'$th string.
Now, one can parametrize the metric considering $'a'$ as constant,
\begin{equation}
ds^2=-f(r)dt^2+\frac{dr^2}{f(r)}+r^2 h_{ij}dx^idx^j,
\end{equation}
where $h_{ij}$ represents the metric on $2$-dimensional boundary.
Now, the solution satisfying Einstein's equation can be expressed as,
\begin{equation}
f(r)=1-a-\frac{2m}{r}+\frac{Q^2}{r^2}+\frac{r^2}{l^2}
\end{equation}
%where, the cosmological constant $\Lambda$ is parametrized as,
%\begin{equation}
%\Lambda=-\frac{n(n-1)}{2l^2}
%\end{equation}
%\subsection{Black hole thermodynamics}
%Now, we briefly investigate the thermodynamics of charged AdS black holes with quark cloud analyzing the intensive thermodynamic parameters.
The mass of black hole which we shall treat as enthalpy of the black hole in extended phase space is given as,
\begin{equation}
M=H=\frac{r_h}{2}-\frac{ar_h}{2}+\frac{Q^2}{2r_h}-\frac{\Lambda r_{h}^3}{6},
\end{equation}
where $r_h$ is the horizon radius of the black hole. The entropy for any static spherically symmetric charged AdS black hole is computed as,
\begin{equation}
S=\frac{A}{4}=\pi r_{h}^2
\end{equation}
The thermodynamic pressure of the black hole is related to the cosmological constant as,
\begin{equation}
P=-\bigg(\frac{\Lambda}{8\pi}\bigg)
\end{equation} 
%Now, the first law of thermodynamics can be given as,
%\begin{equation}
%dM=TdS+\phi dQ+VdP
%\end{equation}
The Hawking temperature $T$ can be expressed as,
\begin{equation}
T=\frac{1-a}{4\sqrt{S}\sqrt{\pi}}+2P\sqrt{\frac{S}{\pi}}-\frac{Q^2\sqrt{\pi}}{4S^{3/2}}
\end{equation}
The thermodynamic volume and potential is given as,
\begin{eqnarray}
V=\bigg(\frac{\partial M}{\partial P}\bigg)_{S,Q}=\frac{4}{3}\pi r_{h}^3, \quad \phi=\bigg(\frac{\partial M}{\partial Q}\bigg)_{S,P}=\frac{Q}{r_h}
\end{eqnarray}
The specific heat at constant volume $C_V =0$ and the specific heat capacity at constant $P$ can be written as,
\begin{eqnarray}
C_{P}=T\bigg(\frac{\partial S}{\partial T}\bigg)_{Q,P}=\frac{16PS^3+2(1-a)S^2-2Q^2S}{8PS^2-(1-a)S+3Q^2} 
\end{eqnarray}
Considering the string cloud parameter $a$, the first law can be written as,
\begin{equation}
dM=TdS+VdP+\phi dQ+\mathcal{A}da
\end{equation}
the parameter conjugate to $a$ is $\mathcal{A}=-r_h^2/2=-S/2\pi.$

\subsection{Critical behaviors}
It is straightforward to compute the equation of state in terms of temperature and volume as,
\begin{equation}\label{P}
	P(T,V)=\frac{6^{2/3}(-1+a)V^{2/3}+4\pi^{2/3}(Q^2+3TV)}{12(6\pi)^{1/3}V^{4/3}}
\end{equation}
On taking $v=2r_h$, as the specific volume in analogy with the actual thermodynamic specific volume, we can rewrite the equation of state as,
\begin{equation}\label{equation of state}
P(T,v)=\frac{T}{v}-\frac{(1-a)}{2\pi v^2}+\frac{2Q^2}{\pi v^4}
\end{equation}
The critical points can be calculated taking into account the following conditions for criticality\cite{Poshteh:2013pba},
\begin{eqnarray}
\bigg(\frac{\partial P}{\partial v}\bigg)_{T=T_c}=0, \quad \bigg(\frac{\partial^2 P}{\partial v^2}\bigg)_{T=T_c}=0
\end{eqnarray}
One can compute the critical temperature, critical volume and critical pressure respectively as,
\begin{eqnarray}
T_c=\frac{\sqrt{6}(1-a)^{3/2}}{18\pi Q}, \quad v_c=\frac{2\sqrt{6}Q}{\sqrt{(1-a)}}, \quad P_c=\frac{(1-a)^2}{96\pi Q^2} \nonumber
\end{eqnarray}
Taking the parameter $a=0$, we obtain the critical thermodynamic variables for charged RN-AdS black holes in extended phase space.
\subsection{Coexistence curve}
\begin{figure}[h!]
       \centering
           \includegraphics[width=2.3in]{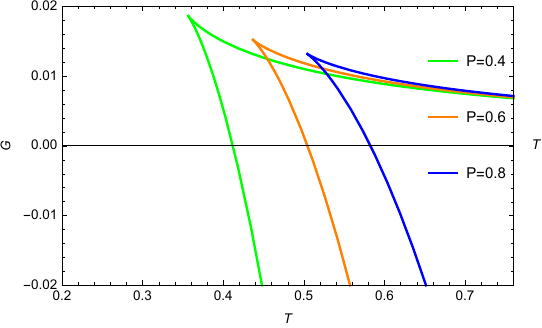}
           \caption{Behavior of Gibbs free energy as a function of $T$ with fixed $a$, $Q$ and $P$.}
           \label{G vs T}
      \end{figure}

\begin{figure}[h!]
       \centering
           \includegraphics[width=2.3in]{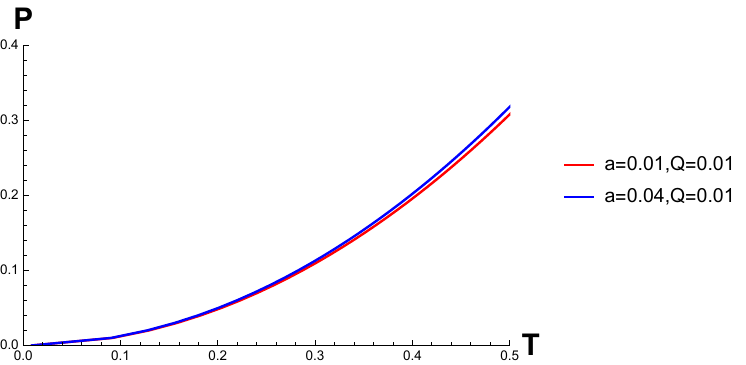}\\
           
           \caption{Behavior of coexistence curve with fixed values of $a$ and $Q$.}
           \label{Coexistence1}
      \end{figure}

The Gibbs free energy for the charged AdS black hole with string cloud can be written as,
\begin{equation}
G(T,Q,P)=M-TS=\frac{(1-a)r_h}{4}+\frac{3Q^2}{4r_h}-\frac{2\pi Pr_{h}^3}{3}
\end{equation}
The behavior of Gibbs free energy is depicted in the below Fig.(\ref{G vs T}).
%We can express the Gibbs free energy as a function of specific volume as,
%\begin{equation}\label{gibbs free energy}
%G=\frac{3Q^2}{2v}+\frac{v(1-a)}{8}-\frac{\pi Pv^3}{12},
%\end{equation}
Utilizing the critical thermodynamic variables, we can write the Gibbs free energy at critical points,
\begin{equation}
G_c=\sqrt{\frac{2(1-a)}{3}}Q
\end{equation}
We can observe the first order phase transition of black hole along the coexistnce curve, so the black hole will have same Gibbs free energy. The equation for coexistence curve is computed as,
\begin{equation}
T=\frac{2\sqrt{P}}{3}\bigg(\frac{6}{\pi}\bigg)^{1/4} \bigg(\sqrt{\frac{6\big(1-a\big)}{\pi}}-8\sqrt{P}Q\bigg)^{1/2}
\end{equation}
The behavior of the coexistence curve with fixed values of $a$ and $Q$ is depicted in the below Fig.(\ref{Coexistence1}).
 
One can note that it resembles RN-AdS black hole taking the parameter $a$ to be zero in the reduced phase space. 

\section{Universal thermodynamics and criticality in the boundary CFT}
In this section, we tried investigating the critical behavior of thermodynamic entities for a 3-dimensional boundary CFT. The 3D boundary CFT is dual to AdS black hole in 4-dimensions which are plugged in a D-dimensional superstring models or M theory models\cite{Witten:1998qj}. The solution of these black holes may be associated with $N$ coincident $(d-2)$ branes moving in such higher dimensional models given by $(D,d,k)$ triplet where $k\sim S^{d+k}$, represents the internal space\cite{Dabholkar:2014wpa}. The compactification of $11$-dimensional M theory on sphere $S^{4+3}$ having radius $l$ gives a $4$-dimensional AdS black hole which descends the Newton constant $G_{4}$ as,
\begin{equation}
G_{4}=\frac{G_{11}}{V(\mathcal{S}^7)}=\frac{G_{11}}{\Omega_{7} l^7}
\end{equation}
where $\Omega_{7}$ is the volume of a unit 7-sphere. The AdS radius $l$ is also related to the brane number $N$ via
\begin{equation}
l^{2(d-1)+k}=2^{-\big(\frac{d(4-d)+3}{2}\big)}\pi ^{7(k+2(d-5))-4}N^{\frac{d-1}{2}}l_{p}^{2(d-1)+k},
\end{equation}
where $l_{p}$ is the planck length. On varying $l$ at fixed $G_d$, the color number $N$ is varied on the field theory side,
\begin{equation}
\frac{1}{4\pi \hbar G_{d}}=\gamma \frac{N^{3/2}}{\mathcal{V}},
\end{equation}
where, $\gamma=2^{\frac{1}{2}((d-4)d-7)}\pi^{14d+7k-75}\Omega_{d+k}\Omega_{d-2}$ is a constant and $\mathcal{V}=\Omega_{d-2}l^{d-2}$ is the spatial volume of the boundary conformal field theory with $(d-1)$ dimensions. Keeping $N^{3/2}$ fixed and varying CFT volume can be equivalently considered as variation of $G_{d}$ or the $d$-dimensional $l_{p}$. Taking this into account, we investigate the thermodynamics for the boundary conformal field theory. The inclusion of variation of $G_d$ in first law renders a novel definition for thermodynamic volume of black hole in the bulk without affecting the CFT on the boundary and a generalized form of Smarr relation was given for the boundary field theory. The black hole mass $M$ can be defined with the internal energy $U$ of the large $N$ Yang-Mills theory at the boundary and the black hole temperature $T$ and entropy $S$ to those of the boundary field theory\cite{Chamblin:1999hg}. Taking into account the condition $f(r_h)=0$ at the event horizon $r_h$, we get
\begin{equation}
M=\frac{\Omega_{2}}{8\pi G_{4}}\bigg(\frac{r_{h}^3}{l^2}+(1-a)r_h+\frac{q^2}{r_{h}}\bigg)
\end{equation}
The entropy of the black hole is given as,
\begin{equation}
S=\frac{A_h}{4\hbar G_{4}}=\frac{\Omega_{2}r_{h}^2}{4\hbar G_{4}}
\end{equation}
Considering the parameter $x=\frac{r_h}{l}$ and $y=\frac{q}{l}$ as dimensionless variables and utilizing the Newton's constant $G_{4}$, the internal energy of the boundary CFT is expressed as,
\begin{equation}
U=M=\frac{\hbar \gamma N^{3/2}}{2l}\bigg(\frac{y^2}{x}+(1-a)x+x^3\bigg)
\end{equation}
The entropy of the boundary CFT can now be given as
\begin{equation}
S=\pi \gamma N^{3/2}x^2
\end{equation}
Keeping $x$ and $y$ as fixed in the limit $\hbar \rightarrow 0$, the internal energy of the CFT vanishes as it is a quantum mechanical quantity, however the mass of black hole is a classical quantity on the AdS side. The dimensionless charge corresponding to $R$-charge in the boundary can further be expressed as,
\begin{equation}
\tilde{Q}=\frac{Ql}{4\pi \hbar G_{4}}=\gamma N^{3/2}y
\end{equation}
%Considering the first law of thermodynamics,
%\begin{equation}
%dU=TdS+\Phi d\tilde{Q}-Pd\mathcal{V}
%\end{equation}
Now, we can express the corresponding thermodynamic variables as,
\begin{eqnarray}
 T&=&\bigg(\frac{\partial U}{\partial S}\bigg)_{\tilde{Q},\mathcal{V}}=\frac{\big\{U,\tilde{Q},\mathcal{V}\big\}_{x,y,l}}{\big\{S,\tilde{Q},\mathcal{V}\big\}_{x,y,l}}\nonumber\\
 &=&\frac{\hbar}{4\pi l}\bigg(3x+\frac{(1-a)}{x}-\frac{y^2}{x^3}\bigg),\\
 \Phi&=&\bigg(\frac{\partial U}{\partial \tilde{Q}}\bigg)_{S,\mathcal{V}}=\frac{\big\{U,S,\mathcal{V}\big\}_{x,y,l}}{\big\{\tilde{Q},S,\mathcal{V}\big\}_{x,y,l}}=\frac{\hbar y}{lx},\\
 P&=&-\bigg(\frac{\partial \mathcal{V}}{\partial U}\bigg)_{S,\tilde{Q}}=\frac{\big\{U,S,\tilde{Q}\big\}_{x,y,l}}{\big\{\mathcal{V},S,\tilde{Q}\big\}_{x,y,l}}\nonumber\\
 &=&\frac{\hbar \gamma N^{3/2}}{4l^3\Omega_{2}}\bigg(\frac{y^2}{x}+(1-a)x+x^3\bigg)
\end{eqnarray}
Here, $T$ is temperature of black hole, $\Phi$ is chemical potential and $P$ is thermodynamic pressure of the black hole respectively. Nambu-bracket notation is used for the above given equations of thermodynamic entities. Considering gauge/gravity duality, one can interpret $T$ as the temperature of the quark-gluon plasma. The pressure is identified with the thermodynamic variable $\mathcal{V}$ which is conjugate to the volume of the CFT leading to $P-\mathcal{V}$ criticality of the boundary CFT. Further, the chemical potential is represented by the variable $\Phi$ which is associated with $R$-current dual to the supersymmetric Yang-Mills theory. The inflation point can be calculated applying the condition,
\begin{equation}
\bigg(\frac{\partial T}{\partial S}\bigg)_{\tilde{Q},\mathcal{V}}=\bigg(\frac{\partial^2 T}{\partial S^2}\bigg)_{\tilde{Q},\mathcal{V}}=0,
\end{equation}
for investigating the phase structure of the boundary CFT. The critical dimensionless variabes are given as,
\begin{eqnarray}
x_c=\sqrt{\frac{(1-a)}{6}}, \quad y_c=\frac{(1-a)}{6}
\end{eqnarray}
Utilizing these critical variables we can evaluate the critical temperature, critical pressure, critical chemical potential and critical charge respectively as given below,
\begin{eqnarray}
T_c&=&\frac{2\hbar}{l\pi}\sqrt{\frac{(1-a)}{6}}, \quad P_c=\frac{(1-a)^{3/2}N^{3/2}\gamma \hbar}{3\sqrt{6}\Omega_{2}l^3}\\
\Phi_c&=&\frac{\hbar}{l}\sqrt{\frac{(1-a)}{6}}, \quad \tilde{Q_c}=\frac{1}{6}(1-a)N^{3/2}\gamma
\end{eqnarray}
The dimensionless parameter $T\mathcal{V}^{1/2}$, may be considered relevant in order to determine the phase transition and so, $T$ and $\mathcal{V}$ are not fixed separately as the boundary field theory is conformal making $T$ and $\mathcal{V}$ as dependent parameters. We can consider charge $\tilde{Q}(y)$ as the control parameter instead of $T$ to investigate the thermodynamics of the given system. Further, one can expect to observe quantum phase transition rather than the thermal phase transition considering the above control parameter. The compressibility factor $\tilde{z}$ for the charged AdS black hole with quark cloud can be calculated as,
\begin{equation}
\tilde{z}=\frac{P_c \mathcal{V}_c}{T_c}=\frac{(1-a)}{6}N^{3/2}\pi \gamma
\end{equation}
The heat capacity is divergent at constant charge for the boundary theory and second order phase transition is observed. The heat capacity can be evaluated as,
\begin{eqnarray}
&C_{\tilde{Q}}&=\tilde{T}\bigg(\frac{\partial S}{\partial \tilde{T}}\bigg)_{\tilde{Q}}=\frac{\tilde{T}\{S,\tilde{Q}\}_{x,y}}{\{\tilde{T},\tilde{Q}\}_{x,y}}\nonumber\\
&=&\frac{2N^{3/2}\pi x^2\gamma\big(3x^4+(1-a)x^2-y^2\big)}{\big(3x^4-(1-a)x^2+3y^2\big)}
\end{eqnarray}

\begin{figure}[h!]
       \centering
           \includegraphics[width=2.0in, height=1.2in]{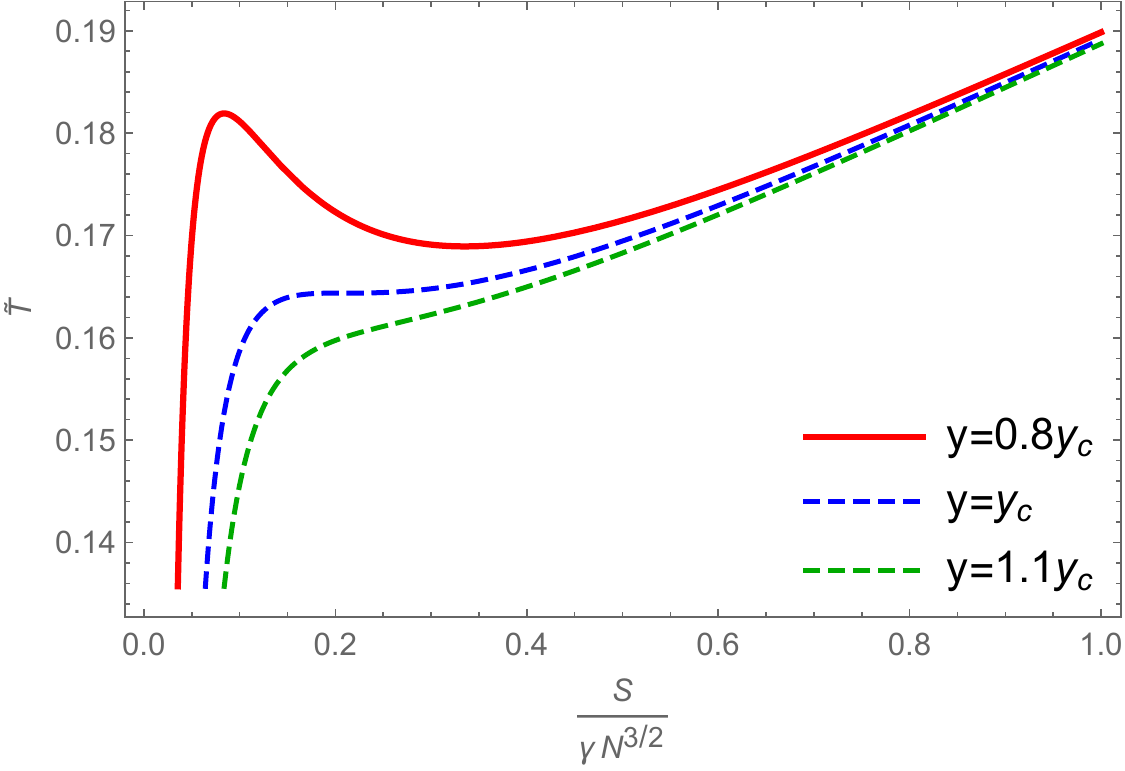}
           \hspace{0.1cm}
           \includegraphics[width=2.1in, height=1.2in]{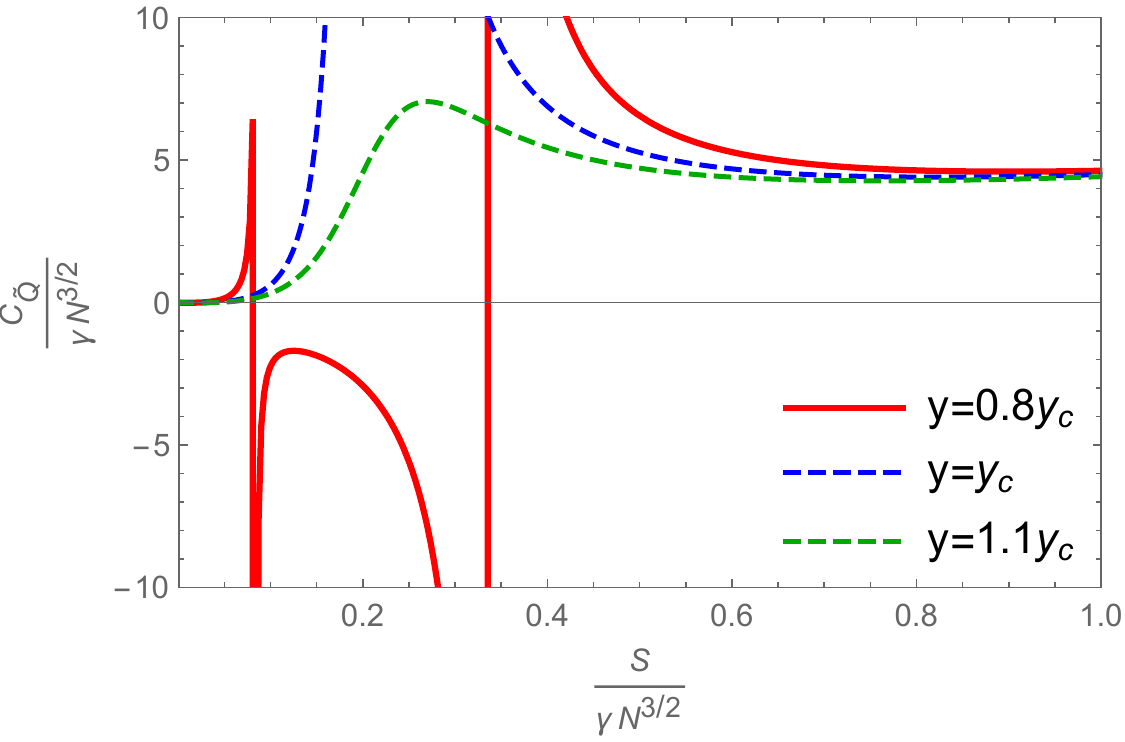}\\
           
          \caption{First figure denotes the behavior of $\tilde{T}$ as a function of $S/\gamma N^{3/2}$. Second figure denotes  the behavior of heat capacity $C_{\tilde{Q}}/\gamma N^{3/2}$ as a function of $S/\gamma N^{3/2}$ for different values of y below and above criticality.}
           \label{TbarvsS}
      \end{figure}
We can observe that the heat capacity in the bulk at fixed value of charge and volume is finite at the critical point for charged AdS black hole with quark cloud but it diverges in the CFT at critical points (see Fig.(\ref{TbarvsS})). With pressure as the defining parameter, we are specifically examining the $P-\mathcal{V}$ criticality of the boundary CFT in this case. We find that the $P-\mathcal{V}$ behavior of the boundary field theory differs somewhat from the behavior of the black hole in the bulk. Two phases appear above critical temperatures, but there is just one phase below it. Additionally, pressure, not volume, jumps across the two-phase regime. If one keeps $\tilde{Q}$ and $\tilde{V}$ fixed, the critical exponents are not mean fields, despite the fact that the CFT equation of state and the van der Waals equation of state are identical. We obtain the critical exponents to be mean field and a phase transition comparable to that of the van der Waals black hole in the bulk when we assume that pressure is the order parameter in the $P-\tilde{V}$ plane. Moreover, it can be noted that critical exponents could transform into mean fields when $\Phi$ is used as the order parameter in the $\tilde{\Phi}-\tilde{Q}$ diagram instead of $\tilde{Q}$. To conjecture phase behavior in terms of $P$ and $\tilde{V}$, one must take into account the dimensionless variables $P/(N^{3/2}T^3)$ and $\mathcal{V}T^{1/2}$. Charge, not temperature, should be used as the control parameter. Additionally, as these variables do not directly scale with temperature, we will determine mean field critical exponents by looking at phase behavior in the $\Phi-\tilde{Q}$ plane.
\par
Now, removing $x$ and $y$ taking $\hat{Q}=Q/Q_c=\tilde{Q}/\tilde{Q_c}$ and $\hat{\Phi}=\Phi/\Phi_c=\tilde{\Phi}/\tilde{\Phi_c}$
we can rewrite the equation of state,
\begin{equation}
T=\frac{3Q^2\hbar^4-N^3\gamma^2\Phi^2(Q^2-(1-a)\hbar^2)}{4N^{3/2}\pi Q\gamma \Phi \hbar^2}
\end{equation}
Again, the critical point may also be obtained from the relation
\begin{equation}
\bigg(\frac{\partial \hat{Q}}{\partial \hat{\Phi}}\bigg)_{\tilde{T}}=\bigg(\frac{\partial^2 \hat{Q}}{\partial \hat{\Phi}^2}\bigg)_{\tilde{T}}=0
\end{equation}
We introduced 
\begin{equation}
\eta=|\hat{\Phi_2}-\hat{\Phi_1|}
\end{equation}
to be the order parameter characterizing the change of phase of the boundary CFT near the critical point analogous to the van der waals system. We notice that there is a remarkable similarity between the phase structure of the boundary field theory and van der waals liquid gas system. A two-phase regime can be noticed in the boundary system when $\tilde{T}>1$.
\par
\begin{figure}[h!]
       \centering
           \includegraphics[width=2.0in, height=1.2in]{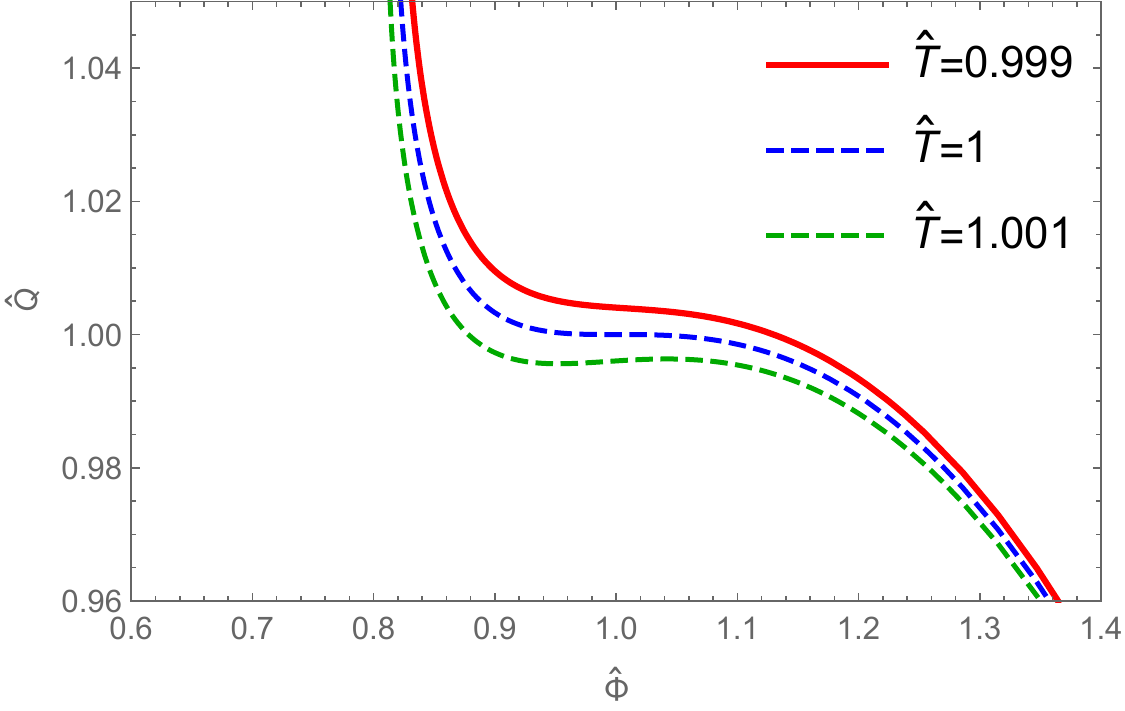}
           \hspace{0.1cm}
           \includegraphics[width=2.0in, height=1.2in]{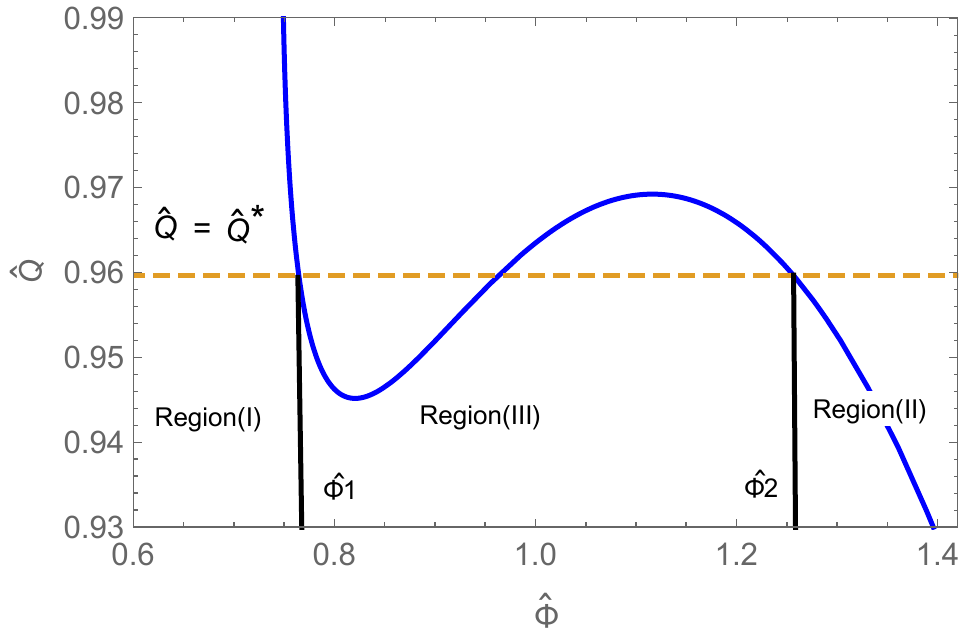}\\
           
          \caption{First figure denotes the behavior of $\hat{Q}$ with respect to $\hat{\Phi}$ for fixed values of $\hat{T}$. Second figure denotes  the behavior of $\hat{Q}$ as a function of $\hat{\Phi}$ above critical value of $T$.}
           \label{Q_phi}
      \end{figure}

Utilizing the Maxwell equal area law, one can substitute the oscillating part of the isotherm shown in Fig.(\ref{Q_phi}) by an isocharge,
\begin{equation}\label{Q}
\int_{\hat{\Phi_1}}^{\hat{\Phi_2}}\hat{Q}d\hat{\Phi}=\hat{Q}*(\hat{\Phi_2})-\hat{Q}*(\hat{\Phi_1})
\end{equation}
where $Q$ is a function of $(\hat{T},\hat{\Phi})$ and is given as,
\begin{equation}
\hat{Q}=\frac{4\hat{T}\hat{\Phi}}{3}-\frac{\hat{\Phi} \sqrt{(-18+16\hat{T}^2+3\hat{\Phi}^2)}}{3}
\end{equation}

\begin{figure}[h!]
       \centering
           \includegraphics[width=2.0in]{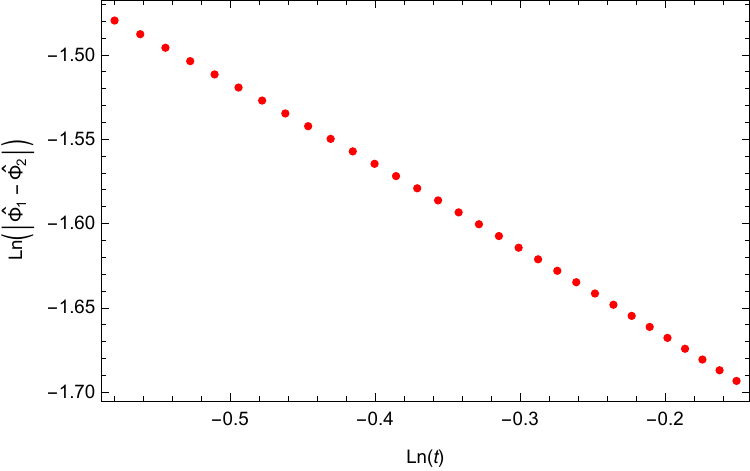}
           \caption{Behavior of $Ln (\hat{\Phi_1}-\hat{\Phi_2})$ vs $Ln(t)$.}
           \label{t_phi}
      \end{figure}

Now, we can characterize critical exponents by defining
\begin{eqnarray}
t=\hat{T}-1, \quad \Phi=\hat{\Phi}-1
\end{eqnarray}
to determine the behavior of physical quantities near the critical point. The behavior of the chemical potential difference $\eta$ can be characterized by the critical exponent $\lambda$ along the coexistence curve given by $\hat{Q}-\hat{T}$ and can be expressed as
\begin{equation}
\eta=|\hat{\Phi}_{1}-\hat{\Phi}_{2}|=bt^{\lambda}
\end{equation}
for $t>0$. The nature of critical exponents is plotted in Fig.(\ref{t_phi}) as a function of $Ln(t)$.

\begin{figure}[h!]
       \centering
           \includegraphics[width=2.0in, height=1.2in]{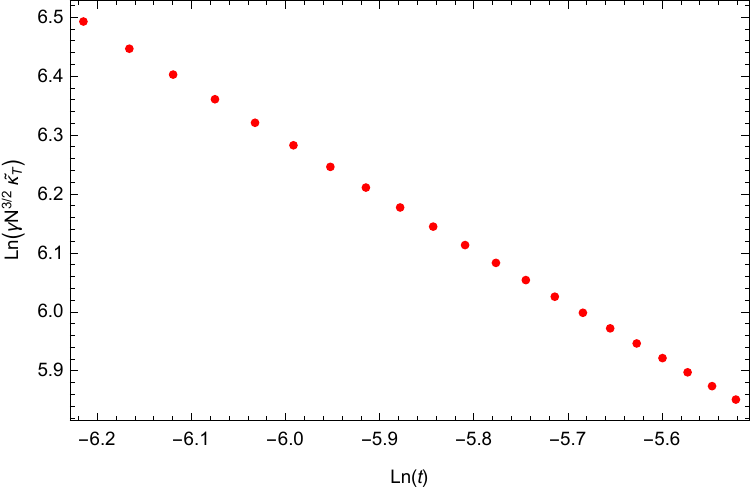}
           \hspace{0.1cm}
           \includegraphics[width=2.1in, height=1.2in]{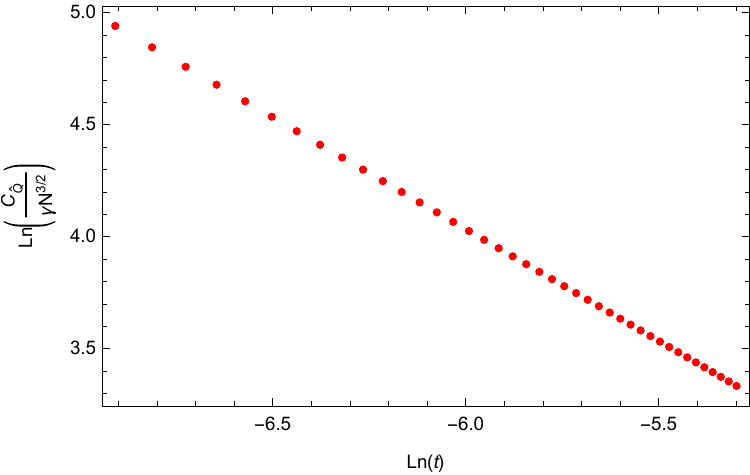}\\
           
          \caption{Behavior of isothermal compressibility and heat capacity at constant $\Phi$ vs $Ln(t)$ respectively.}
           \label{kappa_phi}
      \end{figure}

Numerical solution of eqn.(\ref{Q}) is fitted into a straight line after generating the points for $t>10^{-4}$ and we obtain, $b=-1.76595$ and $\lambda=0.499096$. Considering reduced parameter space, one can express the isothermal compressibility $\kappa_{\tilde{T}}$ and the adiabatic compressibilty $\kappa_{S}$ respectively as,
\begin{eqnarray}
\kappa_{\tilde{T}}=-\frac{1}{\tilde{\Phi}}\bigg(\frac{\partial \tilde{\Phi}}{\partial \tilde{{Q}}}\bigg)_{\tilde{T}}=-\frac{1}{\hat{Q}_c\hat{\Phi}}\bigg(\frac{\partial \hat{Q}}{\partial \hat{\Phi}}\bigg)_{\hat{T}}
\\
\kappa_{\tilde{S}}=-\frac{1}{\tilde{\Phi}}\bigg(\frac{\partial \tilde{\Phi}}{\partial \tilde{{Q}}}\bigg)_{\tilde{S}}=-\frac{1}{\hat{Q}_c\hat{\Phi}}\bigg(\frac{\partial \hat{Q}}{\partial \hat{\Phi}}\bigg)_{\hat{S}}
\end{eqnarray}
The heat capacity at constant chemical potential is given as (see Fig.(\ref{kappa_phi})),
\begin{equation}
C_{\tilde{\Phi}}=\tilde{T}\bigg(\frac{\partial S}{\partial \tilde{T}}\bigg)_{\tilde{\Phi}}
\end{equation}
Utilizing the expression of $C_{\tilde{\Phi}}$, we can obtain the relation between heat capacity and compressibilty as,
\begin{equation}
\kappa_{\tilde{T}}C_{\tilde{\Phi}}\big(\kappa_{S}C_{\tilde{\Phi}}\big)^{-1}=1
\end{equation}
The heat capacity and compressibility posses critical behaviors near the critical point generally,
\begin{equation}
C_{I}= \begin{cases}
t^{-\alpha_I}, \quad t>0,\\
-(t)^{-\alpha_I'}, \quad t<0
\end{cases}
\end{equation}
and
\begin{equation}
\kappa_{I}= \begin{cases}
t^{-\gamma_I}, \quad t>0,\\
-(t)^{-\gamma_I'}, \quad t<0
\end{cases}
\end{equation}
i.e, along the coexistence curve given by $\hat{Q}-\hat{\Phi}$ and isochemical potential curve $\hat{\Phi}=1$. Substituting $\hat{T}=1+(-t)$ in the functions of heat capacity and compressibility along the isochemical potential curve given by $\hat{\Phi}$ and  further expansion to lowest order in $(-t)$, we obtain,
\begin{eqnarray}
\nonumber C_{\tilde{Q}} \approx \frac{(-1+a)N^{3/2}\pi \gamma}{9(-t)}, \quad C_{\tilde{Q}}^c&=&\frac{4\pi \big((-1+a)N^{3/2}\gamma \big)}{3}
\\ \nonumber
\kappa_{\tilde{T}} \approx \frac{1}{2(-1+a)N^{3/2}\gamma (-t)}, \quad
\kappa_{S}^c&=&-\frac{6}{(1-a)N^{3/2}\gamma}
\end{eqnarray}
that determines $\alpha_{\tilde{Q}}'=\gamma_{\tilde{T}}'=1$ and $\alpha_{\tilde{\Phi}}'=\gamma_{\tilde{S}}'=0$ for $t<0$. The negative signature in the adiabatic compressibility $\kappa_{S}$ and the heat capacity $C_{\tilde
{\Phi}}$ signifies the system instability. These kind of instabilities have also been observed for charged AdS black holes and rotating black holes with constant angular momentum.

\section{Thermodynamic geometry}
In extended phase space, the mass of the black hole is regarded as the enthalpy and hence we call it as enthalpy representation. Utilizing enthalpy as the key potential, the behavior of black hole microstructures is speculated in $(T,V)$-plane. The novel equation of state as a function of temperature and volume is considered to compute the Ruppeiner scalar curvature and study the effect of quark cloud parameter $a$ on the microstructures of charged AdS black hole.
%The line element given is the basis for formulating the Ruppeiner geometry. From the first law of thermodynamics, we have;
%\begin{equation}
%dU=TdS-PdV+\sum_{i=1}^{N}\mu_idn_i
%\end{equation}
%where, $n_i$ gives the thermodynamic variables and $\mu_i$ the chemical potentials. Therefore, the line element can be re-written as,
%\begin{equation}
%dl^2=\frac{C_V}{T^2}dT^2+\frac{(\partial_n \mu)_T}{T}dn^2
%\label{line element}
%\end{equation}
%\subsection{Ruppeiner metric and Ruppeiner scalar curvature} 
The Ruppeiner metric\cite{Wei:2015iwa,Wei:2019uqg} is actually defined as the Hessian of the entropy and the line element in the $(T,V)$-plane is given as,
\begin{equation}\label{metric}
dl_{R}^2=\frac{C_V}{T^2}dT^2+\frac{1}{T}\bigg(\frac{\partial P}{\partial V}\bigg)_{T}dV^2
\end{equation}
where, $C_{V}$ is the specific heat at constant volume which is taken to be zero for static black holes. A derivation of the line element is given in\cite{Singh:2020tkf,Singh:2023hit,Singh:2023ufh,Singh:2024msw,Mahish:2020gwg,Ghosh:2019rsu,Yerra:2020oph,Wei:2019ctz}. The Ruppeiner scalar contains the physical information related to the microscopic interactions which is considered as an empirical tool to measure the interaction strength of black hole molecules. It can also signify the type of interaction which dominates among black hole molecules i.e., weather the interaction is attractive or repulsive interaction which depends on the sign of Ruppeiner scalar curvature.
\par
\begin{figure}[h!]
       \centering
           \includegraphics[width=2.2in]{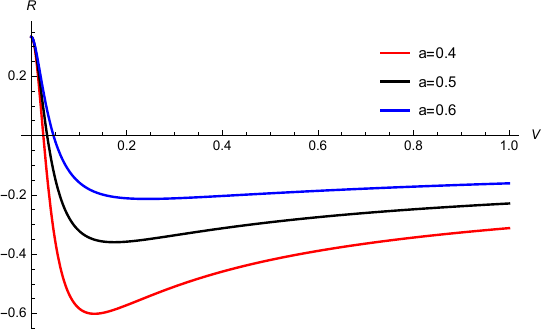}
           
         \caption{Behavior of scalar curvatur with fixed values of $a$ and $Q$.}
                     \label{R vs V}
      \end{figure}

Utilizing the equation of state given in eqn.(\ref{P}), we can evaluate the corresponding Ruppeiner scalar curvature in different planes. Since $C_{V}$ is zero here, we can fix $C_{V}$ to be a small non zero value for computing $R$ and $C_{V}$ comes to be an overall constant factor multiplicative to it. Considering the $(T,V)$-plane we can write the scalar curvature as,
\begin{eqnarray}\label{RTV}
\nonumber R_{TV}&=&\frac{\big(3(-1+a)V^{2/3}+2\times 6^{1/3} \pi^{2/3}(2Q^2+3TV)\big)}{3\big(3(-1+a)V^{2/3}+6^{1/3} \pi^{2/3}(4Q^2+3TV)\big)^2}\\
&\times& \big(4\times 6^{1/3} \pi^{2/3}Q^2+3(-1+a)V^{2/3}\big)
\end{eqnarray}
The Ruppeiner scalar $R_{TV}$ is plotted in Fig.(\ref{R vs V}) as function of $V$ for fixed values of string cloud parameter $a$ and black hole charge $Q$.

Here, one can note that for extremal limits i.e., $T=0$, the Ruppeiner scalar curvature $R_{TV}=1/2$ signaling the dominance of repulsive type interaction among black hole microstructures and the scalar curvature is positive. For smaller values of $a$ the strength of interaction increases and with increasing $a$ it decreaeses. For vanishing $a$ the behavior of Ruppeiner scalar curvature comes out to be similar to the case of RN-AdS black hole.

\subsection{Thermodynamic curvature criticality in the boundary CFT}
\begin{figure}[h!]
       \centering
           \includegraphics[width=2.1in]{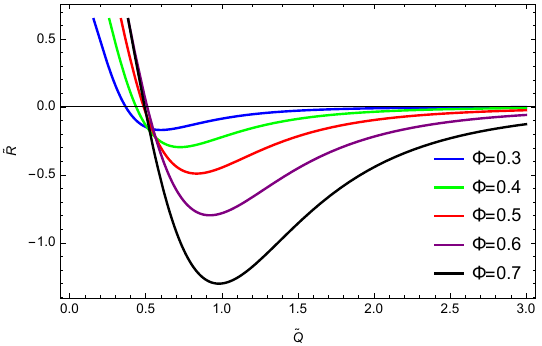}
          \caption{Behavior of $\tilde{R}$ with $\tilde{Q}$ for fixed $a, \gamma, \tilde{\Phi}$ and  $\hat{N}$ .}
                     \label{RQ vs Q}
      \end{figure}

Weinhold\cite{Weinhold:1975xej} defined Reimannian metric as the Hessian of the internal energy $M$ and took derivatives with respect to entropy and extensive thermodynamic variables. These two metrics were found to be conformally equivalent\cite{Rafiee:2021hyj,Cong:2021jgb,HosseiniMansoori:2020jrx,Ahmed:2023snm,Ahmed:2023dnh,HosseiniMansoori:2024bfi} having $\beta$ as the conformal factor which is the inverse of temperature,
\begin{equation}
ds_R^2=\beta ds_W^2
\end{equation}
Here, in this section we see that the Ruppeiner geometry which was diverging at the point where the heat capacity was diverging in the bulk is now unable to achieve a one-to-one correspondence between phase transition points at the boundary CFT and the singularity of the scalar curvature at the boundary. This result is anamalous to the results obtained in the bulk but one can remove the anamoly by defining some new formulation of thermodynamic phase space geometry. However, as it is not successful to give appropriate explanation for the thermal stability of a thermodynamic system at certain critical points, we are not going into detailed formalism of this. At the boundary, isochemical potential $\tilde{\Phi}$ and charge $Q$ is considered as appropriate fluctuation coordinates and the scalar curvature is calculated in $(\tilde{Q}, \tilde{\Phi})$-plane for the charged AdS black holes with quark cloud. The heat capacity at constant charge is given as,
\begin{equation}
C_{\tilde{Q}}=\frac{(-1+a)\hat{N}^{3/2}\pi \gamma \tilde{Q}^2\big(-3\tilde{Q}^2-6\tilde{\Phi}^2+\tilde{\Phi}^4 \big)}{9\tilde{\Phi}^2\big(\tilde{Q}^2-2\tilde{\Phi}^2+\tilde{\Phi}^4\big)}
\end{equation}
The scalar curvature in the $(\tilde{Q}, \tilde{\Phi})$-plane is computed as,
\begin{eqnarray}
\tilde{R}&=&\frac{R_A}{R_B} \\ \nonumber R_{A}&=&-648\tilde{\Phi}^6\big(9\tilde{Q}^6+\tilde{Q}^2\tilde{\Phi}^2(30-7\tilde{\Phi}^2)\\
\nonumber &-&3\tilde{Q}^4\tilde{\Phi}^2(-9+\tilde{\Phi}^2)+\tilde{\Phi}^6(-6+\tilde{\Phi}^2)^2(-1+\tilde{\Phi}^2)\big)\\
\nonumber	R_{B}&=&(-1+a)\hat{N}^{3/2}\pi \gamma (-3\tilde{Q}^2-6\tilde{\Phi}^2+\tilde{\Phi}^4)\big(9\tilde{Q}^4\\ \nonumber
&+&\tilde{\Phi}^4(-6+\tilde{\Phi}^2)^2-18\tilde{Q}^2\tilde{\Phi}^2(-2+\tilde{\Phi}^2)\big)^2
\end{eqnarray}
and the behavior is shown in Fig.(\ref{RQ vs Q}).

\section{Remarks}
The fact that cosmological constant is regarded as a thermodynamic variable, it is quite natural to question how bulk pressure or volume on the boundary CFT should be interpreted. In boundary field theory, the variation of pressure, or $\Lambda$, is identical to the variation of color number $N$. The conjugate of pressure is thermodynamic volume, which in boundary field theory can be understood as the chemical potential $\mu$ for color. The spatial volume of the field theory can also be varied by considering the color number $N$ to be fixed and just varying $\Lambda$ in the bulk. Using $\Phi$ as the order parameter, we calculated critical exponents for heat capacities in the $\hat{Q}-\hat{\Phi}$ plane while accounting for the latter interpretation. If one keeps $\tilde{Q}$ and $\tilde{V}$ fixed, the critical exponents are not mean fields, despite the fact that the CFT equation of state and the van der Waals equation of state are identical. We obtain the critical exponents to be mean field and a phase transition comparable to that of the van der Waals black hole in the bulk if we consider that pressure is the order parameter in the $P-\tilde{V}$ plane. Moreover, it can be seen that critical exponents may turn into mean fields when $\Phi$ is used as the order parameter in the $\tilde{\Phi}-\tilde{Q}$ plane instead of $\tilde{Q}$.
\par 
Further, we investigated the thermodynamic curvature in the bulk and  criticality at the boundary conformal field theory. The scalar curvature is dependent on the string cloud parameter $a$, and the strength of interaction among the black hole molecules changes as the string cloud density changes. For vanishing $a$, the behavior of Ruppeiner scalar curvature comes out to be similar to the case of RN-AdS black hole. For the boundary CFT one can note that the Ruppeiner geometry which was diverging at the point where the heat capacity was diverging in the bulk is now unable to achieve a one-to-one correspondence between phase transition points at the boundary CFT and the singularity of the scalar curvature at the boundary. This result is anamalous to the results obtained in the bulk but one can remove the anamoly by defining some new formulation of thermodynamic phase space geometry. Probing the nature of $R$ is useful for obtaining preliminary understanding of the dominant interactions (at least empirically) at a specific temperature, horizon radius or in a specific parameter regime for black holes.
\section*{Acknowledgements}
A.S. is thankful to CSIR for financial support under project no. 03WS(003)/2023‑24/EMR‑II/ASPIRE. A.S. would also like to thank Indian Institute of Technology, Bhubhaneswar where bulk of this work was completed as a graduate student there. We are extremely thankful to Chandrasekhar Bhamidipati for insightful discussions.
   
\end{document}